\documentstyle[aps,prl,floats,psfig]{revtex} 
\begin{document}
\firstfigfalse \firsttabfalse


\title{Atomic structure and vibrational properties of icosahedral
B$_4$C boron carbide}

\author{R. Lazzari,$^1$\cite{stgobain} N. Vast,$^1$\cite{e-nathalie}
J.M. Besson,$^2$ S. Baroni ,$^{3,4}$\cite{e-stefano} and A. Dal
Corso$^5$\cite{andrea}}

\address{ $^1$ CEA, Centre d'Etudes de Bruy\`eres, 91680
Bruy\`eres Le Ch\^atel, France\\ $^2$Physique des Milieux Condens\'es,
UMR CNRS 7602, B77, Universit\'e P. et M. Curie, 4, Place Jussieu,
75252 Paris, France \\ $^3$ Scuola Internazionale Superiore di Studi
Avanzati and INFM, Via Beirut 2-4, 34014 Trieste, Italy \\ $^4$ Centre
Europ\'een de Calcul Atomique et Moleculaire, ENS-Lyon, 46 All\'ee
d'Italie, 69007 Lyon, France \\ $^5$ Institut Romand de Recherche
Num\'erique en Physique des Mat\'eriaux, EPFL, Ecublens, 1015
Lausanne, Switzerland}

\date{\today}

\maketitle

\begin{abstract} The atomic structure of icosahedral B$_4$C boron
carbide is determined by comparing existing infra-red absorption and
Raman diffusion measurements with the predictions of accurate {\it ab
initio} lattice-dynamical calculations performed for different
structural models, a task presently beyond X-ray and neutron diffraction ability. By examining
the inter- and intra-icosahedral contributions to the stiffness we show
that, contrary to recent conjectures, intra-icosahedral bonds are
harder. \end{abstract}


\narrowtext

Covalently bonded solids based on boron, carbon, or nitrogen 
form the hardest materials  presently known \cite{louie}, and   $\rm
B_4C$ comes third after diamond and cubic BN, with the advantages of
being easily synthesized and  stable up to very high
temperatures \cite{superhard}. Hence it is  used as abrasive or shielding material sustaining  extreme conditions, while $^{10}B$-enhanced
ceramics are used in  nuclear reactors.

\begin{figure} [htbp] $$\psfig{file=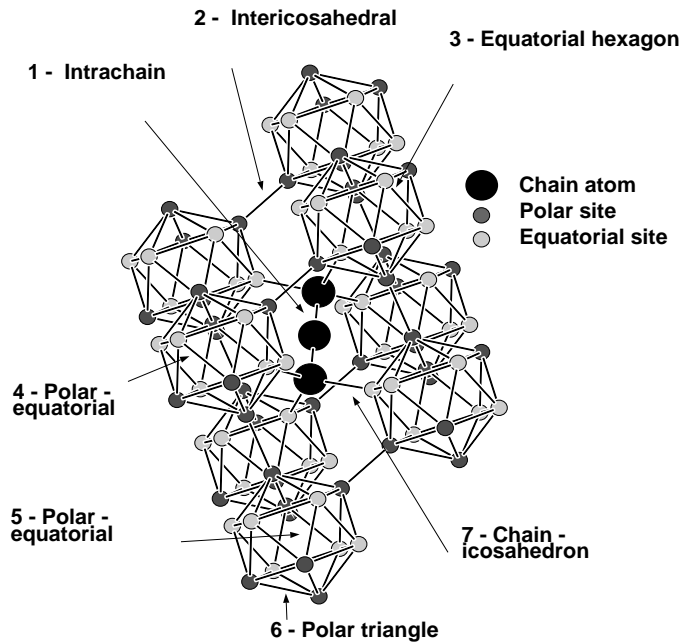,width=8cm}$$ \small
\caption{Atomic structure of B$_4$C.} \label{fig:balls} \end{figure}

The atomic structure responsible for these properties is rather unique
\cite{donohue}: an arrangement of distorted $\rm B_{11}C$ icosahedra
located  at the nodes of a rhombohedral  Bravais lattice
($R \bar 3 m $ space group, Fig. \ref{fig:balls}). This is a
modification of the $\alpha$-$\rm B_{12}$ structure, which
accommodates three-atom chains along the (111) rhombohedral axis,
linking different B$_{11}$C icosahedra. 
 Two inequivalent crystallographic sites exist in the
icosahedron. The six atoms which form the top and bottom triangular
faces of the icosahedron sit at the {\it polar} sites, and are
directly linked to atoms in neighboring icosahedra; the other six
corners of the icosahedron form a puckered hexagon in the plane
perpendicular to the [111] axis, and their symmetry-equivalent sites
are called {\it equatorial}. Each one of the six equatorial atoms is
linked to an inter-icosahedral chain.

An experimental determination of the atomic structure of $\rm B_4C$ is
still lacking \cite{WerhII,jimenez}. Neutron diffraction
\cite{neutron,neutron2} cannot distinguish $^{11}$B from $^{12}$C
because their  scattering lengths are too close
\cite{neutron}. X-ray diffraction has allowed to identify a C-B-C
chain \cite{X,larson}, but the location of the remaining C atom in the
B$_{11}$C icosahedron remains unsettled because the X-ray form factors
of boron and carbon atoms are also too close. A recent attempt to
resolve the atomic structure of B$_4$C using photo-emission and X-ray
absorption spectra \cite{jimenez} did not yield conclusive results.

In this paper the determination of the atomic structure of B$_4$C is
addressed by a combination of first-principles calculations, based on
density-functional theory (DFT) and density-functional perturbation
theory (DFPT), with infra-red absorption and Raman diffusion
results. We have considered three ordered
configurations of B$_4$C: in the {\it chain} configuration, all the
icosahedra are entirely constituted by boron atoms---B$_{12}$---while
all the carbon atoms lie entirely in the inter-icosahedral
chains---C--C--C; in the {\it polar} configuration,  the chains are
 assumed to be C--B--C, in agreement with X-ray
diffraction data \cite{X,larson}, while one of the polar atoms
of the icosahedra is substituted by a carbon atom---B$_{11}$C; the
{\it equatorial} configuration is similar to the polar one, but 
in this case  the substitution involves  an equatorial atom.
  These configurations are highly idealized, and some degree
of substitutional disorder is expected to occur in actual 
samples. Nevertheless, their vibrational properties are sufficiently
different so as to allow one to discriminate between different
structural models by comparing the observed Raman and infra-red
spectra with the predictions of accurate lattice-dynamical
calculations. We finally determine the equation of
state of B$_4$C  and point out that the recently postulated {\it inverted
molecular behavior} \cite{aim} is not consistent
with our theoretical findings.

Calculations were performed within DFT and DFPT \cite{dfpt}, using the
local-density approximation and the plane-wave pseudo-potential
method. The pseudo-potentials of boron and carbon  were respectively
the same as in Ref. \cite{vast} and \cite{martintrou}. Plane-waves up
to a kinetic energy cutoff of 65 Ry have been included in the basis
set. The irreducible wedge of the Brillouin zone has been sampled with
10 and 2 points for static and dynamic properties respectively in the
chain model, and 20 and 3 points in the polar and equatorial
ones. The structural parameters reported in table I have been obtained by
minimizing the crystal energy with respect to the size, shape  and
 internal degrees of freedom of
the unit cell. The {\it
chain} model has $R \bar 3 m $ symmetry, while the
substitution of one B atom in the icosahedron induces a small
monoclinic distortion, which, for the polar and
equatorial configurations, amounts to 1.8\% and 0.5\% of the
rhombohedral cell length respectively, and to 1\% and 0.1\% of the
rhombohedral angle respectively. The calculated data agree well with
X-ray \cite{X} and neutron diffraction \cite{neutron} data (see
Table I) for all of the three configurations. An  agreement
better than 3\%  is found for almost all of the bond lengths,  which
 depend very little on the configuration. Exceptions to this
are the intra-chain bond (bond \#1 in Fig. \ref{fig:balls}) whose
length would be underestimated by 10\% in the chain model and the
chain-icosahedron bond (bond \# 7 in Fig. \ref{fig:balls}) which is
predicted to be 5\% too short both in the polar and the equatorial
models. We interpret the failure to accurately predict the
chain-icosahedron bond as due to our neglect of disorder effects in
the chain. Occasionally, the  chain could be
B--C--C\cite{neutron2}, a possibility which we did not consider. The
calculated energies  are  also very close: those of 
B$_{12}$-CCC and equatorial B$_4$C are slightly higher by
respectively 73 and 4 meV/atom as compared to polar B$_4$C which would
then result to be the ground state. However, our neglect of disorder
and  of finite-temperature effects does not allow us to draw any
definite conclusions.

\begin{table} 
\caption{Comparison between the structural parameters of B$_4$C as
calculated for three structural models ({\it Chain}, {\it Polar}, {\it
Equatorial}: see text) with those observed experimentally.  $a$ is the
lattice parameter (\AA), $\alpha$ is the angle between the
rhombohedral lattice vectors (degrees), while $b_n$ indicates the
length of the $n$-th bond (\AA) (see Fig. \ref{fig:balls}). }
\vskip8pt

\begin{tabular}[t]{lcccrr} 
 & {\it Chain} & {\it Polar} & {\it Equatorial} &Expt.\tablenotemark[1] &
Expt.\tablenotemark[2]  \\ 
\hline
$a$ & 5.12 & 5.10  & 5.13  & 5.163 & 5.155 \\
$ \alpha $ & 65.9  & 65.8  & 64.9  & 65.732
& 65.679 \\ 
$b_1$ & 1.30 & 1.42 & 1.43 & 1.434 & 1.438 \\
$b_2$ & 1.70 & 1.71 & 1.69 & 1.716 & 1.699 \\ 
$b_3$ & 1.73 & 1.73 & 1.72 & 1.693 & 1.687 \\
$b_4$ & 1.77 & 1.78 & 1.78 & 1.758 & 1.760 \\
$b_5$ & 1.77 & 1.77 & 1.77 & 1.762 & 1.761 \\
$b_6$ & 1.81 & 1.78 & 1.78 & 1.805 & 1.810 \\
$b_7$ & 1.65 & 1.59 & 1.59 & 1.675 & 1.669 \\
\end{tabular} 
\tablenotetext[1] {Ref.  \protect\cite{X};} 
\tablenotetext[2] {Ref.   \protect\cite{neutron}.} 
\end{table}

\begin{figure} [htbp] \small\def\baselinestretch{1.2}
$$\psfig{file=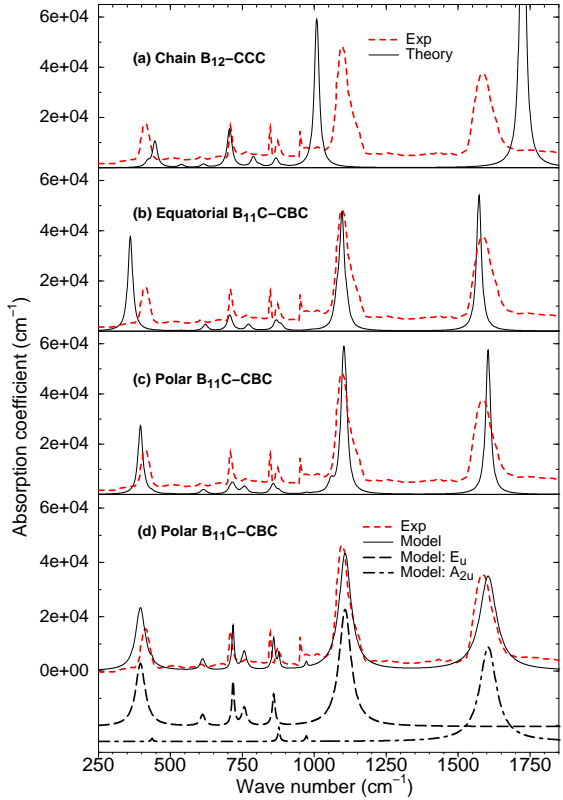,width=8cm}\quad\quad\quad $$ \caption{Panels a-c:
infrared spectra of B$_4$C as calculated for the three configurations
considered in this work (solid lines) compared with that deduced from
experiments [\protect{\ref{WerhI}}] (dashed lines). (a) B$_{12}$-CCC;
(b) equatorial B$_4$C; (c) polar B$_4$C. Panel d: calculated
spectrum of polar B$_4$C fitted to account for the experimental mixing
of polarization. $E_u$ (dash) and $A_{2u}$ (dot-dash) contributions
shifted for clarity. } \label{fig:infrared} \end{figure}

In order to substantiate the hypothesis of a   polar B$_4$C, we have decided to
compare the vibrational spectrum predicted using state-of-the-art
theoretical methods (DFPT) with that observed experimentally. As the
difference of the atomic masses of boron and carbon amounts to 11$\%$,
one expects to observe a difference on those vibrational modes
which involve mainly one of the two sites. In panels a-c of
Fig.~\ref{fig:infrared} we display the infrared spectrum as calculated
for the three configurations and as deduced from reflectivity
experiments \cite{WerhI}. The polarization-averaged absorption
coefficient reads \cite{pruesch}: \begin{equation} \alpha(\omega)= {2
\pi^2 \over 3 c n_1 \Omega} \sum_j \sum_\alpha \Big | \sum_{\beta s}
Z^{*s}_{\alpha \beta}\ \ e_{\beta s}^j \Big |^2 \delta(
\omega-\omega_j),
\label{eq:abs-coeff} \end{equation} where $Z^{*s}_{\alpha\beta}$ is
the Born effective-charge tensor of $s$-th atom, $c$ is the speed of
light, $\Omega$ is the unit cell volume, $\alpha$ and $\beta$ are
Cartesian coordinates, $j$ labels the vibrational mode, and
$\omega_j$, $e_{\beta s}^j$ the corresponding eigen-frequency and
displacement pattern, respectively. For the polar and equatorial
configurations, the dynamical matrix has been averaged with respect to
all the symmetry-equivalent substitution sites in the icosahedron, so
that rhombohedral symmetry is recovered also in these cases, and $E_u$
($A_{2u}$) modes are infrared active when light is polarized
perpendicular  (parallel) to the [111] axis.  Phonon
life-time and other broadening effects have been empirically accounted
for by replacing the $\delta$ functions in Eq. (\ref{eq:abs-coeff})
with Lorentzians with a half width of 10 cm$^{-1}$.  The {\it positions} 
of the peaks agree only 
for  the polar configuration.

The agreement between the calculated and observed peak {\it intensities} is
still rather poor, mainly due to our lack of knowledge of the
individual line-widths as well as of the polarization conditions of
the experiment with respect to the orientation of the sample. 
A simple adjustment of the theoretical data turns out to
bring them in  good agreement with experiments. We
have re-weighted all the $A_{2u}$ peaks with respect to the $E_{u}$
ones by a same factor which is treated as a fitting
parameter. Once this is done, we have kept  the   integrated intensity 
 of  each mode at its theoretical value, 
and we have   utilized the widths of individual
peaks as free parameters. The resulting values of the fitting
parameters are reported in table II. The agreement between the
experimental spectrum and the one predicted for the polar
configuration is now very good (Fig. \ref{fig:infrared} d), and only
the peak at 950 cm$^{-1}$ is missed by the theory. Its shape is asymmetric and its intensity varies from sample to sample
\cite{WerhI,stein}, hence suggesting that it is not a simple bulk
lattice vibration. In fact, theory predicts an infrared mode at 972
cm$^{-1}$ but its small intensity should hardly be observable (Table
II). It might be activated by surface effects and/or bulk
defects. Actually experiments are performed on hot-pressed
micro-crystalline samples, where the defect density is very high.
This leads to a degradation of the reflectivity spectrum with respect
to the ideal (mono-crystalline) case, and the largest experimental
error is expected at the LO minima.

\begin{table} \caption{Polar B$_4$C: theoretical infrared active
frequency $\omega$ (cm$^{-1}$) and relative integrated intensity
$I_{rel}$. The highest intensity of each polarization is 
normalized to 10. $ \Gamma$ is the half width of the Lorentzians used
in Fig. 2d (cm$^{-1}$).} \begin{center} \begin{minipage}{8.6cm}
\begin{tabular}{llccccccc}
$E_u$ &$\omega$ & 396 & 519 & 616 & 714 & 761 & 857 & 1103 \\
 &$I_{rel}$ & 4.1 & $\approx 0.$ & 0.3& 0.6 & 0.4 & 0.6 & 10. \\
 &$\Gamma$ & 15. & 5. & 10. & 5. & 10. & 3. & 20. \\
$A_{2u}$& $\omega$ & 436 & 720 & 877 & 972 & 1606 & & \\ 
 &$I_{rel}$ & 0.1 & $\approx 0.$ & 0.2 & 0.1 & 10. & & \\
 &$\Gamma $ & 4. & 4. & 4. & 4. & 30& & \\
 \end{tabular}
\end{minipage}
\end{center}
\end{table}

\begin{figure} [htbp] \small\def\baselinestretch{1.2}
$$\psfig{file=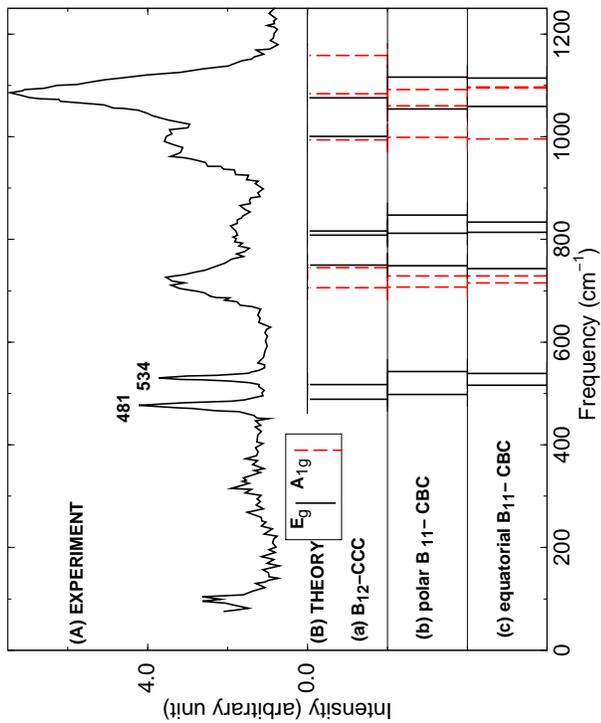,width=8cm}\quad\quad$$ \caption{B$_4$C Raman
spectrum: (A) experiment \protect\cite{tallant} (B) theory for (a)
B$_{12}$-CCC, (b) polar B$_4$C, (c) equatorial B$_4$C. Solid lines:
$E_g$ mode; dashed lines: $A_{1g}$ mode.} \label{fig:raman}
\end{figure}

In order to further confirm  the  polar configuration
as  the dominant one  in B$_4$C, we now compare the Raman
spectrum measured on a single crystal \cite{tallant}
with the predictions of our calculations (Fig.~\ref{fig:raman}).
Unfortunately, a complete quantitative comparison is difficult due to
mode broadening, especially at high frequency. However, two modes at
481 and 534 cm$^{-1}$ are rather sharp \cite{tallant}, and they are
properly accounted for only by the polar model, which predicts two
peaks respectively at 498 and 543 cm$^{-1}$.  The former
is a  rotation of the chain about an axis perpendicular to
the [111] direction  and the later is the
librational mode of the icosahedron previously identified in
$\alpha$-boron \cite{vast}.  These two modes have similar frequencies
and widths, and they both involve angular atomic displacements,
so that the errors made in the prediction of their positions are
expected to be similar and to cancel to a large extent.  The distance
of 54 cm$^{-1}$ between the two peaks is therefore a stringent test
for the assignment, and only the polar configuration does account for
it, while it is underestimated by a factor of two in the equatorial
model (Fig.~\ref{fig:raman}). At higher frequencies,
structural disorder is likely to be responsible for most of the
observed broadening. Twinning or stacking faults--- recently
 identified as pairs of neighboring twin defects---are observed in
single crystals \cite{zuppi}. They distort the crystal only slightly
because they preserve the rhombohedral symmetry and the structure of
the icosahedron. In B$_4$C ceramics obtained by hot pressing, however,
these defects can be as close as a few lattice parameters
\cite{zuppi}, and the low-frequency part of the Raman spectrum is
quite different from those of Fig. 3. 
In particular, while no
Raman activity is expected in clean samples below 360 cm$^{-1}$, a
broad band systematically appears at low frequency in ceramics
\cite{tallant,simeone}, twice-peaked around 250 and 320 cm$^{-1}$,
with an intensity which differs from sample to sample of a same
bulk composition \cite{tallant}. We attribute this band to a high
density of states of disorder-activated acoustic phonons, in analogy
to what is found, {\it e.g.}, in ice VII \cite{besson}.

\begin{table}
\caption{Theoretical and experimental bulk modulus (GPa) of the unit
cell and of the icosahedron. Theoretical pressure derivatives are
found to be: $B_0^\prime$=3.5.} \begin{tabular}[t]{lcccccc} & & This work
 & X-ray & Neutron & Ultrasound \\ \hline
$B_{4}C$ & Unit cell & 248 & 245\tablenotemark[1] & 220\tablenotemark[2] &
 247\tablenotemark[3] \\ & icosahedron & 274 & - &
169\tablenotemark[2] & - \\ $\alpha$-boron & Unit cell & 222 &
224\tablenotemark[4] & & \\ & icosahedron & 297 & - & & 
 
\end{tabular}
\tablenotetext[1] {Ref. \protect\cite{loveday};}
\tablenotetext[2] {Ref. \protect\cite{aim};}
\tablenotetext[3] {Ref. \protect\cite{Bb4c-aselage};}
\tablenotetext[4] {Ref. \protect\cite{bbore}.}
\end{table}

Having shown that the vibrational  spectra of B$_4$C can be
consistently interpreted in terms of a carbon atom occupying one of
the polar sites, we  consider now its  elastic properties,
 which have recently attracted some interest
\cite{aim}. Theoretical  volume  {\it vs.} pressure  data - determined up
to a an hydrostatic pressure of 10 GPa - have been  fitted to a
Murnaghan's equation of state: the resulting bulk modulus $B_0$ is
reported in Table III together with the `compressibility of the
icosahedron' evaluated as in Ref. \cite{aim}. Our theoretical value is
close both to ultrasonic measurements \cite{Bb4c-aselage} and to
unpublished X-ray data \cite{loveday}. No evidence of the {\it
inverted molecular behavior} reported in Ref. \cite{aim} was found
from our calculations. In Ref. \cite{aim}, a powder sample of NaCl and
B$_4$C was used. Since the two materials have widely different elastic
constants, they might have experienced very  different
 pressures if, as it is often the case, some
strain is transmitted at the grain boundaries (Voigt model). As no
correction  for this effect was made,
the bulk modulus was probably found  too low since at a given
NaCl pressure (calibrant) the actual pressure experienced by B$_4$C
was higher. For this reason the apparent bulk modulus of the
icosahedra, which was found smaller than B$_0$, might  need
to be verified in a quasi hydrostatic environment, with an accurate
measurement of pressure. Moreover,  twinning in boron allows the 
crystal to accommodate large strains
and to prevent the formation of quasi-crystalline phases
\cite{zuppi}. We are  inclined to attribute the observed
anti-molecular compression to pressure inhomogeneities induced by the
intensive twinning in B$_4$C ceramics. An experiment on a single
crystal would provide a definite answer to this issue.

In conclusion, the atomic structure of B$_4$C consists of one
icosahedron B$_{11}$C with the carbon atom staying at a polar site,
and a chain C-B-C. B$_4$C  and  $\alpha$ boron are not molecular
nor---as we have demonstrated---{\it inverted molecular}
crystals. Rather, they should be considered as members of a new class
of covalently bonded materials.

The authors are grateful to J. Loveday and D. Simeone for the
communication of unpublished results. N.V. thanks L. Zuppiroli for a
fruitful discussion.


\begin{references}

\bibitem[*] {stgobain} Presently at Laboratoire du Verre et des
Interfaces, CNRS-SAINT GOBAIN, 93303 Aubervilliers, France.

\bibitem[**] {e-nathalie} Author to whom correspondence should be
sent. Electronic address: vast@bruyeres.cea.fr .

\bibitem[+]{e-stefano} Electronic address: baroni@sissa.it .

\bibitem[\dagger]{andrea} Presently at  SISSA, Trieste,
Italy. dalcorso@sissa.it . 


\bibitem{louie} 
H. T. Hall and L.A. Compton, Inorg. Chem. {\bf 4}, 1213 (1965);
S. Han, J. Ihm, S.G. Louie and M.L. Cohen, Phys. Rev. Lett. 
{\bf 80}, 997 (1998).

\bibitem{superhard} I.J. McColm, {\it Ceramic Hardness}, Plenum press,
New York (1990).


\bibitem{donohue} J. Donohue, {\it The Structure of the Elements},
John Wiley and Sons, New York (1974).

\bibitem{WerhII} H. Werheit, U. Kuhlmann and T. Lundstr\"{o}m,
Jour. of Alloys and Compounds {\bf 204}, 197 (1994).

\bibitem{jimenez} I. Jim\'enez, D.G.J. Sutherland, T. van Buuren,
J.A. Carlisle, L.J. Terminello and F.J. Himpsel, Phys. Rev. B {\bf
57}, 13167 (1998).

\bibitem{neutron} B. Morosin B., G.H. Kwei, A.C. Lawson, T.L. Aselage
and D. Emin, Jour. of Alloys and Compounds {\bf 226}, 121 (1995).

\bibitem{neutron2} G.H. Kwei and B. Morosin, Jour. of Physical
Chemistry {\bf 100}, 8031 (1996).

\bibitem{X} B. Morosin, T.L. Aselage and R.S. Feigelson,
Mat. Res. Soc. Symp. Proc. {\bf 97 }, 145 (1987).

\bibitem{larson} A.C. Larson, {\it Boron Rich Solids. Conf. Proc. {\bf
140} }, AIP, New York (1986), p. 109.

\bibitem{aim} R.J. Nelmes, J.S. Lodevay, R.M. Wilson, W.G. Marshall,
J.M. Besson, S. Klotz, G. Hamel, T.L. Aselage T.L. and S. Hull,
Phys. Rev. Lett. {\bf 74}, 2268 (1995).



\bibitem{dfpt} S. Baroni, P. Giannozzi, and A. Testa, Phys. Rev. Lett.
{\bf 58}, 1861 (1987); P. Giannozzi, S. de Gironcoli, P. Pavone, and
S. Baroni, Phys. Rev. B {\bf 43}, 7231 (1991).

\bibitem{vast} N. Vast, S. Baroni, G. Zerah, J.M. Besson, A. Polian,
J.C. Chervin and T. Grimsditch, Phys. Rev. Lett. {\bf 78}, 693
(1997).

\bibitem{martintrou} N. Troullier and J.L. Martins, Phys. Rev. B {\bf
43}, 1993 (1990).
 
\bibitem{WerhI} U. Kuhlmann, H. Werheit and K.A. Schwetz, Jour. of
Alloys and Compounds {\bf 189}, 249 (1992). The
absorption index has been obtained from the measured reflectivity
spectrum using the Kramers-Kr\"onig relation. From this we have then
estimated the absorption coefficient by setting the real part of the
refraction index to $n_1 = 2.7$, as its dependence on frequency is not
known to us. This procedure does not affect the positions of the
peaks and provides an estimate of the absorption coefficient
within 10 to 20$\%$ which is precise enough for our
purposes. \label{WerhI}

\bibitem{pruesch} P. Br\"uesch, {\it Phonons: Theory and Experiments
II }, p. 50, Springer Verlag, Berlin (1986).

\bibitem{stein} H. Stein, T.L. Aselage and D. Emin,
Ref. \cite{Bb4c-aselage}, p. 322.

\bibitem{tallant} D.R. Tallant, T.L. Aselage, A.N. Campbell and
D. Emin, Phys. Rev. B {\bf 44}, 2535 (1991).

\bibitem{zuppi} P. Favia, T. Stoto, M. Carrard, P.A. Stadelmann and
L. Zuppiroli, Micros. Microanal. Microstruc. {\bf 7}, 225 (1996).

\bibitem{simeone} D. Simeone, private communication (1997).

\bibitem{besson} J.M. Besson, M. Kobayashi, T. Nakai, S. Endo and
Ph. Pruzan, Phys. Rev. B {\bf 55}, 11191 (1997). 

\bibitem{Bb4c-aselage} J.H. Gieske, T.L. Aselage and D.Emin, {\it
Boron Rich Solids. Conf. Proc. {\bf 231}}, AIP, New York (1991),
p. 377.

\bibitem{loveday} J. Loveday, private communication (1997).

\bibitem{bbore} R.J. Nelmes, J.S. Lodevay, D.R. Allan, J.M. Besson,
G. Hamel, S. Klotz, P. Grima and S. Hull Phys. Rev. B {\bf 47}, 7668
(1993).

\end{references}
\end{document}